\documentclass[letterpaper,11pt]{article}
\usepackage[hmargin=1in,vmargin=1in]{geometry}
\usepackage{amsmath}
\usepackage{graphicx}
\usepackage{bm}
\usepackage{epsfig}
\usepackage{xcolor}
\title{On the initial conditions of scalar and tensor fluctuations in $f(R,\phi)$ gravity}
\author{ S. Cheraghchi$^1$, F. Shojai$^{1,2}$\\$^1$Department of Physics, University of Tehran,\\ Tehran, Iran.\\$^2$Foundations of Physics Group, School of Physics,\\ Institute for Research in Fundamental Sciences (IPM),\\ Tehran, Iran.\\}
\begin{document}
\maketitle
\begin{abstract}
We have considered the perturbation equations governing the growth of fluctuations in generalized scalar
tensor theory during inflation. we have found that the scalar metric perturbations at very early times are negligible compared with the scalar field perturbation, just like general relativity. At sufficiently early times, when $q/a\gg H$, we have obtained the metric and scalar field perturbation in the form of WKB solutions  up to an undetermined coefficient. Then we have quantized the scalar fluctuations and expanded the metric and the scalar field perturbations with the help of annihilation and creation operators of the scalar field perturbation.  The standard commutation relations of annihilation and creation operators  fix the unknown coefficient. Going over to the gauge invariant quantities which are conserved beyond the horizon, we have obtained the initial condition of the generalized Mukhanov-Sasaki equation. And a similar procedure is performed for the case of tensor metric perturbation.
\end{abstract}
\section{Introduction}
The structure formation at the early universe is one of the most important issue in modern cosmology \cite{Weinberg,Dodel,Lyth}. The large scale structure formation is explained by the gravitational instabilities of the space-time metric and matter. To study these instabilities, it is necessary to eliminate the gauge ambiguities coming from the freedom in the coordinate choice in general relativity. To do this, usually a specific gauge is firstly chosen,  gauge fixing,  and then the results are expressed in terms of gauge invariant variables which represent the physical quantities.

On the other hand, the inflationary mechanism provides the initial condition for cosmic perturbations observed in the cosmic background radiation (CMB) today. The gravitational instability is amplified during inflationary era and at the end of it, the curvature perturbation, remains frozen at superhorizon scales. It is a gauge invariant quantity which provides the seed of galaxy formation at the time of horizon crossing during the radiation dominated era. Thus, we need to work out the equations of motion for the gauge invariant quantities,  introduced by the scalar and tensorial perturbations of metric and the matter field. The initial conditions of these equations can be found by considering the fluctuations at sufficiently early time of inflation when the perturbations are deep inside the horizon. At these very early times, the perturbations are essentially quantum fields which can be expanded in terms of creation and annihilation operators satisfying in the standard commutation relations. Moreover, it is usually assumed that the initial quantum state of the inflaton field is the standard vacuum state\footnote{Other choices of the initial state of quantum inflaton include the $\alpha$-vacua \cite{Danielsson:2002kx}, the coherent state \cite{Kundu:2011sg,Kundu:2013gha}, the $\alpha$-states \cite{Mottola:1984ar,Allen:1985ux}, the thermal state \cite{Bhattacharya:2005wn}, and the excited-de Sitter modes \cite{Yusofi:2014mta}.}, Bunch-Davies vacuum. Putting all these points together, one can find the initial conditions of the classical equations governing the evolution of perturbations at inflationary era.

Here we are interested in deriving the initial conditions  of modified Mukhanov-Sasaki equation in generalized scalar tensor gravity,  described by the Lagrangian density $\sqrt{-g}f(R,\phi)$ \cite{living, Faraoni, Mukhanov92, Capozziello}. Many studies of primordial perturbations in this theory was firstly formulated by Hwang and his collaborators \cite{Hwang1, Gong, Hwang2, Hwang3}.
They have applied the conformal equivalence of generalized gravity theories with general relativity minimally coupled to a scalar field, in the absence of ordinary matter. Then they have studied the evolution of the curvature perturbation considering the second order action.  Here we shall use a straightforward calculations based on using some appropriate gauges and after obtaining the generalized Mukhanov-Sasaki equation, we shall find its initial condition. 

The organization of this paper is as follows: In the next section, we will consider the scalar metric perturbations and also the scalar field fluctuations. After deriving the equations governing the dynamics of these perturbations, these equations are simplified imposing the Newtonian gauge condition. Then we have paid special attention to these equations at sufficiently early times, when $q/a\gg H$, and used WKB approximation. This specifies solutions to an unknown coefficient. To find it, we quantize the scalar field perturbations and demand that the creation and annihilation operators satisfy the standard commutation relations. In section 3, by introducing a special gauge, the generalized Mokhanov-Sasaki equation is obtained.  The initial condition of this equation is given using the results of the previous section. Finally in the last section, we  have followed a similar procedure to find the initial conditions for tensor metric perturbation. 
\section{Scalar perturbations in generalized scalar-tensor gravity}

We consider inflation driven by a single scalar field which is coupled non-minimally with gravity and represented by the following action: 
\begin{equation}\label{action1}
S=\int \sqrt{-g}\left(\frac{1}{2 \kappa^2}f(R,\phi)-\frac{1}{2}\omega(\phi)g^{\alpha\beta}\partial_\alpha\phi\partial_\beta\phi-V(\phi)\right)~d^4x.
\end{equation}
where $\kappa^2=8\pi G$. Varying above action with respect to the metric and scalar field,  we obtain:

\begin{equation}\label{b1}
3FH^2=\frac{1}{2}(RF-f)-3H\dot F+\kappa^2\left\lbrack\frac{1}{2}\omega\dot\phi^2+V(\phi)\right\rbrack
\end{equation}
\begin{equation}\label{b2}
-2F\dot H=\ddot F-H\dot F+\kappa^2\omega\dot\phi^2
\end{equation}
\begin{equation}\label{b3}
\ddot\phi+3H\dot\phi+\frac{1}{2\omega}\left(\omega_{,\phi}\dot\phi^2+2V_{,\phi}-\frac{f_{,\phi}}{\kappa^2}\right)=0
\end{equation}
where $F(R,\phi)\equiv\frac{\partial f}{\partial R}$, $H$ is Hubble parameter, $R=6(2H^2+\dot{H})$    is the Ricci scalar. 

Perturbing the scalar field as $\phi=\phi_0+\delta\phi$ and take the line element of perturbed FRW universe as:
\begin{equation}\label{metric}
ds^2=-(1+2\Phi)dt^2-2a(t) \partial_i\beta dtdx^i+a^2(t)(\delta_{ij}-2\Psi\delta_{ij}+2\partial_i\partial_j\gamma+\mathcal{D}_{ij})dx^idx^j.
\end{equation}
where $\Phi$, $\beta$, $\Psi$ and $\gamma$ are scalar perturbations and $\mathcal{D}_{ij}$ is tensor perturbation which is the traceless, divergenceless and symmetric. Under these conditions, we can see that the corresponding Fourier modes, $\mathcal{D}_{ij}(q,t)$, have two independent components and can be expressed in terms of the polarization tensors, $\hat{e}^+_{ij}$ and $\hat{e}^\times_{ij}$, as following 
\begin{align}\label{condition}
\mathcal{D}_{ij}(q,t)=\hat{e}^+_{ij}\mathcal{D}_+(q,t)+\hat{e}^\times_{ij} \mathcal{D}_\times(q,t)\hspace{0.5in}e_{xx}^+=-e_{yy}^+=1,\hspace{0.2in}e_{xy}^\times=e_{yx}^\times=1
\end{align}
where the momentum $q$ is considered along the $z$-axis. We will back to the above relation in Section 4.
\\
Now, perturbing the background field equations (\ref{b1})-(\ref{b3}) at linear order for scalar perturbations gives \cite{living}:
\begin{eqnarray}\label{p1}
&-\frac{\Delta}{a^2}\Psi+H A=-\frac{1}{2F}\left[ \left(3H^2+3\dot H+\frac{\Delta}{a^2}\right)\delta F
-3H\dot{\delta F}\right.\nonumber
\\
& \left. +\frac{1}{2}\left(\kappa^2\omega_{,\phi}{\dot\phi_0}^2+2\kappa^2V_{,\phi}-f_{,\phi}\right)
\delta\phi+\kappa^2\omega\dot\phi_0{\dot{\delta\phi}}+\left(3H\dot F-\kappa^2\omega\dot\phi_0^2\right)\Phi+\dot F A\right]
\end{eqnarray}
\begin{equation}\label{p2}
H\Phi+\dot\Psi=\frac{1}{2F}\left\lbrack\kappa^2\omega\dot\phi_0\delta\phi+\dot{\delta F}-H\delta F-\dot F \Phi\right\rbrack
\end{equation}
\begin{equation}\label{chi}
\dot\chi+H\chi-\Phi+\Psi=\frac{1}{F}\left(\delta F-\dot F\chi\right)
\end{equation}
\begin{eqnarray}\label{p3}
&\dot A+ 2H A+\left(3H+\frac{\Delta}{a^2}\right)\Phi=\frac{1}{2F}\left[ 3\ddot{\delta{ F}}+3H\dot{\delta{F}}-\left(6H^2+\frac{\Delta}{a^2}\right)\delta F+4\kappa^2\omega\dot\phi_0\dot{\delta{\phi}}\right.\nonumber \\
&\left. +(2\kappa^2\omega_{,\phi}\dot\phi_0^2-2\kappa^2V_{,\phi}+f_{,\phi})\delta\phi-3\dot F\dot \Phi-\dot F A-(4\kappa^2 \omega\dot\phi _0^2+3H\dot F+6\ddot F)\Phi\right]
\end{eqnarray}
\begin{equation}\label{p4}
\begin{split}
&\ddot{\delta{F}}+3H\dot{\delta{F}}-\left(\frac{\Delta}{a^2}+\frac{R}{3}\right)\delta F+\frac{2}{3}\kappa^2\omega\dot{\phi_0}\dot{\delta{\phi}}+\frac{1}{3}(\kappa^2\omega_{,\phi}\dot{\phi_0^2}-4\kappa^2 V_{,\phi}+2f_{,\phi})\delta\phi=\\
&\dot F(A+\dot\Phi)+\left(2\ddot F+3H\dot F+\frac{2}{3}\kappa^2\omega\dot{\phi_0^2}\right)\Phi-\frac{1}{3}F\delta R
\end{split}
\end{equation}
\begin{equation}\label{p5}
\begin{split}
&\ddot{\delta\phi}+\left(3H+\frac{\omega_{,\phi}}{\omega}\dot\phi_0\right)\delta\dot\phi+\left\lbrack-\frac{\Delta}{a^2}+\left(\frac{\omega_{,\phi}}{\omega}\right)_{,\phi}\frac{\dot{\phi_0^2}}{2}+\left(\frac{2V_{,\phi}-\frac{f_{,\phi}}{\kappa^2}}{2\omega}\right)_{,\phi}\right\rbrack\delta\phi\\
&=\dot\phi_0\dot\Phi+\left(2\ddot\phi_0+3H\dot\phi_0+\frac{\omega_{,\phi}}{\omega}\dot{\phi_0^2}\right)\Phi+\dot\phi_0 A+\frac{1}{2\omega\kappa^2}F_{,\phi}\delta R
\end{split}
\end{equation}
where  
\begin{equation}\label{eq1}
\chi\equiv a(\beta+a\dot\gamma)\hspace{0.5in}
A\equiv3(H\Phi+\dot\Psi)-\frac{\Delta}{a^2}\chi
\end{equation}
and the Ricci perturbation is given by:
\begin{equation}\label{sp}
\delta R=-2\left\lbrack\dot A+4H A+\left(\frac{\Delta}{a^2}+3\dot H\right)\Phi-2\frac{\Delta}{a^2}\Psi\right\rbrack.
\end{equation}
Working in Fourier space significantly simplifies calculations. Thus we write the perturbation equations in Fourier space and choose the Newtonian gauge. It is a particularly simple gauge to use for the scalar mode of metric perturbations and  does not leave a residual gauge symmetry. In this gauge, $\gamma=\beta=0$ and thus $(\ref{eq1})$ yields:
\begin{equation}\label{sade}
\chi= 0\hspace{0.5in}
A=3(H\Phi+\dot\Psi).
\end{equation}
Denoting the perturbations of metric and the scalar field by their corresponding Fourier transforms:
\begin{align}
\Phi(\textbf x,t)=\int{\frac{d^3q}{(2\pi)^{3/2}}\Phi_q(t)e^{-i\textbf q.\textbf x}}.
\end{align}
\begin{align}
\Psi(\textbf x,t)=\int{\frac{d^3q}{(2\pi)^{3/2}}\Psi_q(t)e^{-i\textbf q.\textbf x}}.
\end{align}
\begin{align}
\delta\phi(\textbf x,t)=\int{\frac{d^3q}{(2\pi)^{3/2}}\delta\phi_q(t)e^{-i\textbf q.\textbf x}}.
\end{align}
We insert these relation and (\ref{sade}) into equations (\ref{p1})-(\ref{p5}), and get the following
equations for the Fourier modes:
\begin{equation}\label{eq2}
\begin{split}
&\left\lbrack3\left(H^2-\dot H+3H\frac{\dot F}{F}\right)+\frac{q^2}{a^2}-\frac{\kappa^2\omega\dot\phi_0^2}{F}\right\rbrack\Phi_q+\left\lbrack3\left(H^2+\dot H-H\frac{\dot F}{F}\right)+\frac{q^2}{a^2}\right\rbrack\Psi_q\\
&+3\left(H+\frac{\dot F}{F}\right)\dot\Psi_q+3H\dot\Phi_q
=-\frac{\kappa^2\omega\dot\phi_0}{F}\delta\dot\phi_q-\frac{\kappa^2}{2F}\left(\omega_{,\phi}\dot\phi_0^2+2V_{,\phi}-\frac{f_{,\phi}}{\kappa^2}\right)\delta\phi_q,
\end{split}
\end{equation}
\begin{equation}\label{eq3}
\left(H+2\frac{\dot{F}}{F}\right)\Phi_q+\dot\Phi_q+\left(H-\frac{\dot{F}}{F}\right)\Psi_q+\dot\Psi_q-\frac{\kappa^2\omega\dot\phi_0}{F}\delta\phi_q=0
\end{equation}
\begin{align}\label{eq4}
\begin{split}
&\ddot\Psi_q+\ddot\Phi_q-\left(\frac{\dot F}{F}-3H\right)\dot\Psi_q+3\left(\frac{\dot F}{F}+H\right)\dot\Phi_q\\
&-\left[\frac{\ddot F}{F}+H\frac{\dot F}{F}-\frac{1}{3}\left(6H^2-\frac{q^2}{a^2}\right)\right]\Psi_q\\
&+\left[2\left(H+\dot H-H^2+H\frac{\dot F}{F}\right)-\frac{q^2}{3a^2}+\frac{1}{3F}(4\kappa^2 \omega\dot\phi _0^2+3H\dot F+6\ddot F)\right]\Phi_q\\
&=\frac{8\kappa^2\omega\dot\phi_0}{3}\dot{\delta\phi_q}+\frac{2}{3}(2\kappa^2\omega_{,\phi}\dot\phi_0^2-2\kappa^2V_{,\phi}+f_{,\phi})\delta\phi_q
\end{split}
\end{align}
\begin{align}\label{eq5}
\begin{split}
&\ddot\Psi_q+\ddot\Phi_q+\left(\frac{\dot F}{F}+5H\right)\dot\Psi_q+\left(3\frac{\dot F}{F}+5H\right)\dot\Phi_q+\left(\frac{q^2}{3a^2}+\frac{R}{3}-3H\frac{\dot F}{F}-\frac{\ddot F}{F}\right)\Psi_q\\
&+\left[\frac{q^2}{3a^2}-\frac{R}{3}+3\frac{\ddot F}{F}+8H^2+4\dot H+9H\frac{\dot F}{F}+\frac{2}{3F}\kappa^2\omega\dot\phi_0^2\right]\Phi_q\\
&=\frac{2}{3}\kappa^2\omega\dot\phi_0\dot{\delta\phi}+\frac{1}{3}(\kappa^2\omega_{,\phi}\dot{\phi_0^2}-4\kappa^2 V_{,\phi}+2f_{,\phi})\delta\phi
\end{split}
\end{align}
\begin{equation}\label{eq6}
\begin{split}
&\delta{\ddot\phi}_q+\left(3H+\frac{\omega_{,\phi}}{\omega}\dot\phi_0\right)\delta\dot\phi_q+\left\lbrack\frac{q^2}{a^2}+(\frac{\omega_{,\phi}}{\omega})_{,\phi}\frac{\dot{\phi_0^2}}{2}+(\frac{2V_{,\phi}-\frac{f_{,\phi}}{\kappa^2}}{2\omega})_{,\phi}\right\rbrack\delta\phi_q\\
&=\left\lbrack2\ddot\phi_0+6H\dot\phi_0+\frac{\omega_{,\phi}}{\omega}\dot{\phi_0^2}-\frac{F_{,\phi}}{\omega\kappa^2}\left(6\dot H+12H^2-\frac{q^2}{a^2}\right)\right\rbrack\Phi_q+\left(\dot\phi_0-3H\frac{F_{,\phi}}{\kappa^2\omega}\right)\dot\Phi_q\\
&+3\left(\dot\phi_0-4H\frac{F_{,\phi}}{\kappa^2\omega}\right)\dot\Psi_q-2\frac{q^2F_{,\phi}}{a^2\omega\kappa^2}\Psi_q-3\frac{F_{,\phi}}{\omega\kappa^2}\ddot\Psi_q
\end{split}
\end{equation}
where in equations $(\ref{eq2})-(\ref{eq6})$, $\delta F$ and $\delta R$ are eliminated by using the following relations
\begin{equation}\label{eq6'}
-\Phi_q+\Psi_q=\frac{\delta F}{F}
\end{equation}
\begin{align}\label{eq6''}
\delta R=-2\left\lbrack\left(12H^2+6\dot H-\frac{q^2}{a^2}\right)\Phi_q+3H\dot\Phi_q+2\frac{q^2}{a^2}\Psi_q+3\ddot\Psi_q+12H\dot\Psi_q\right\rbrack.
\end{align}
Equations (\ref{eq4}), (\ref{eq5}) and (\ref{eq6}) are the dynamical equations of motion for $\Phi$, $\Psi$ and $\delta\phi$ and two remaining equations, (\ref{eq2}) and (\ref{eq3}), are two constraints imposed on 
the fluctuations. This is similar to general relativity in which there are two dynamical equations for the  metric and scalar fluctuations together with a constraint equation in the Newtonian gauge\footnote{See equations (10.1.12)-(10.1.14) from \cite{Weinberg}} .

Now the above perturbation equations, can be used to find the functions $\Psi_q$, $\Phi_q$, $\delta\phi_q$ at sufficiently early times by WKB approximation. At these very early times, all perturbations are inside the horizon and the perturbation modes oscillate more quickly than the expansion rate of the universe, i.e. $q/a\gg H$. The general solutions in the WKB approximation reads:
\begin{align}\label{gs}
\begin{split}
&\Psi_q(t)\longrightarrow g(t)\exp\left(-iq\int_{t^*}^{t}\frac{dt'}{a(t')}\right)\\
&\Phi_q(t)\longrightarrow h(t)\exp\left(-iq\int_{t^*}^{t}\frac{dt'}{a(t')}\right)\\
&\delta\phi_q(t)\longrightarrow y(t)\exp\left(-iq\int_{t^*}^{t}\frac{dt'}{a(t')}\right)
\end{split}
\end{align}
where the rate of change of $g(t)$, $h(t)$ and $y(t)$ are much smaller than $q/a$ and $t^*$ is an arbitrary time. Also since the fluctuation of fields are real functions, the complex conjugate of any above solution, is another independent solution. Substituting $(\ref{gs})$ into $(\ref{eq2})-(\ref{eq6})$ and working up to leading order in $q/a$, one quickly finds that the terms in $(\ref{eq6})$ of second order in $q/a$ cancel each other. The other equations are satisfied if:
\begin{align}\label{eq7}
\frac{g+h}{y}=i\frac{a\kappa^2\omega\dot\phi_0}{q F}
\end{align}
This equation shows that at sufficiently early times, the metric fluctuations  are one order smaller than the scalar fluctuations and thus they will be insignificant. This is a useful result where we see that it  holds not only for the scalar field minimally coupled to gravity in general relativity but also in $f(R,\phi)$ theory. To first order in $q/a$, equation $(\ref{eq6})$ leads:
\begin{align}\label{eq9}
\dot y+\left(H+\frac{\omega_{,\phi}}{2\omega}\dot\phi_0\right)y=i\frac{qF_{,\phi}}{2a\omega\kappa^2}(g+h)
\end{align}
This equation gives the time dependence of scalar field perturbations. Setting $F=\omega=1$ and using $(\ref{eq6})$, reduces the  equations $(\ref{eq7})$ and $(\ref{eq9})$ to:
\begin{equation}
g=h \hspace{0.5in}\frac{g}{y}=i\frac{a\kappa^2\dot\phi_0}{2q}
\end{equation}
which are agree with those are found in GR \cite{Weinberg}.

From $(\ref{eq7})$ and $(\ref{eq9})$ we see immediately that
\begin{align}\label{eq11}
\dot y+Hy+\frac{\dot\phi_0}{2}\left(\frac{\omega_{,\phi}}{\omega}+\frac{F_{,\phi}}{F}\right)y=0
\end{align}
So
\begin{align}\label{eq12}
y(t)=\frac{C}{a}\frac{1}{\sqrt{F\omega}}
\end{align}
where $C$ is an integration constant and needs to be determined.

As explained above, we have considered here, with the very early universe in which the fluctuations of inflaton field which driving  the inflationary expansion of the universe, have quantum nature . According to (\ref{eq7}), ignoring the metric perturbation, we quantize the scalar field in the unperturbed FRW background such as general relativity. Expanding the scalar field perturbation in terms of two independent solutions, we have:
\begin{align}\label{form1}
\delta\phi(\textbf x,t)=\int{\frac{C}{a}\frac{1}{\sqrt{F\omega}}\left\lbrack
{\alpha(\textbf q)e^{i\textbf q.\textbf x}\exp\left(-iq\int\frac{dt}{a}\right)+\alpha^*(\textbf q) e^{-i\textbf q.\textbf x}\exp\left(iq\int\frac{dt}{a}\right)}\right\rbrack}~\frac{d^3q}{(2\pi)^{3/2}}
 \end{align}
where $\alpha$ and $\alpha^*$ are normalized annihilation and creation operators and we imply that they obey the standard commutation relations:
\begin{align}\label{stan}
\left\lbrack\alpha(\textbf q),\alpha(\textbf q')\right\rbrack=0,\hspace{0.5in}\left\lbrack\alpha(\textbf q),\alpha^*(\textbf q')\right\rbrack=\delta^3(\textbf q-\textbf q')
\end{align}
This gives:
\begin{align}
\left\lbrack\delta\phi(\textbf x,t),\delta\dot{\phi}(\textbf y,t)\right\rbrack=i\frac{2q(2\pi)^3C^2}{Fa^3\omega}\delta^3(\textbf x-\textbf y)
\end{align}
Thus: $C=\frac{1}{(2\pi)^{3/2}\sqrt{2q}}$. This means that the initial conditions of metric and scalar field fluctuations, for $a\longrightarrow 0$, are:
\begin{equation}
\Psi_q(t)+\Phi_q(t)\longrightarrow \frac{i\kappa^2\dot{\phi_0}}{(2\pi)^{3/2}}\sqrt{\frac{\omega}{2q^3 F^3}}\exp\left(-iq\int_{t^*}^{t}\frac{dt'}{a(t')}\right)
\label{Psi}
\end{equation}
\begin{equation}
\delta\phi_q(t)\longrightarrow \frac{1}{(2\pi)^{3/2}a\sqrt{2qF\omega}}\exp\left(-iq\int_{t^*}^{t}\frac{dt'}{a(t')}\right)
\label{phi}
\end{equation}

\section{Mukhanov-Sasaki equation}
In $f(R,\phi)$ gravity, there exists a number of gauge invariant quantities \cite{living}. Here we are interested in three of these variables which are defined in the presence of matter fields, as: 
\begin{equation}
\mathcal{R}=\Psi+\frac{H}{\rho+P}\delta q,\hspace{0.5in}\mathcal{R}_{\delta\phi}=\Psi-\frac{H}{\dot\phi}\delta\phi,\hspace{0.5in}\mathcal{R}_{\delta F}=\Psi-\frac{H}{\dot F}\delta F
\label{R}
\end{equation}
where $\delta q = −(\rho + P )v$. $\rho $, $P$ are the energy density and pressure of the matter field and $v$ characterizes the velocity potential of it.  To have a better description of the above quantities, let us point that in the absence of matter fields, $\mathcal{R}$ which is the comoving curvature perturbation on the uniform-field hypersurface, is equal to $\Psi$. For a single field with a potential $V(\phi)$, using $\rho=\dot \phi/2 + V(\phi)$ and $P=\dot \phi/2 - V(\phi)$, we have $\delta q = − \dot \phi \delta \phi $ and thus $\mathcal{R}_{\delta\phi}$ is identical to $\mathcal{R}$. Also since by a special conformal transformation together with introducing $\sqrt{\frac{3}{2\kappa}}\ln
{F}$ as a new scalar field (discussed in detail in \cite{Faraoni}), one can bring the action (\ref{action1}) into the Einstein form, called Einstein frame, thus $\mathcal{R}_{\delta F}$ is identical to $\mathcal{R}_{\delta\phi}$. 

To study the scalar perturbations generated during inflation, it is useful to derive the evolution equation for the curvature perturbation which is known as the Mukhanov-Sasaki equation.
This equation can be easily derived by an appropriate choice of gauge. As mentioned above, in $f(R,\phi)$ gravity, there exists much more gauge invariant quantities in comparison with general relativity. This has led to more different choices for an appropriate gauge. A convenient gauge is spatially-flat (or uniform-curvature) gauge fixed by the conditions $\Psi=\gamma=0$. Taking this as the gauge condition, the authors of \cite{Hwang1}, derive the Mukhanov-Sasaki equation for $\mathcal{R}_{\delta\phi}$ in the Einstein frame. Then the resulted equation can be mapped back into the first frame (Jordan frame) by the inverse conformal transformation mentioned above. In \cite{living}, the authors focus on the coupling of the form of $f(\phi)R$ and choose the gauge $\delta\phi=\delta F=0$ in which $\mathcal{R}=\mathcal{R}_{\delta\phi}=\mathcal{R}_{\delta F}$. They have pointed out that this gauge can not be applied for a general $f(R,\phi)$ gravity including the non linear terms in curvature\footnote {This is clear, since $\delta F=F_{\phi}\delta \phi+F_{R}\delta R$ and the gauge selection must be compatible with it.}.

Here we concentrate on a general $f(R,\phi)$ theory. Looking at (\ref{R}), we choose a different gauge defined by \footnote{The first condition of (\ref{p}) can be automatically satisfied for some specific $f(R,\phi)$ gravity, for example $f(R)$ gravity with $\phi=0$ and also $f(\phi)R$. For these theories, the 
Mukhanov-Sasaki equation is derived in the uniform curvature gauge in \cite{Hwang1}.}:
\begin{align}
\frac{\delta F}{\dot{F}}=\frac{\delta\phi}{\dot{\phi_0}},\hspace{0.5in}\Psi=0
\label{p}
\end{align}
The main advantage of this gauge is that $\mathcal{R}=0$ and there is a scalar curvature defied as $\tilde{\mathcal{R}}\equiv\mathcal{R}_{\delta F}=\mathcal{R}_{\delta \phi}=-\frac{H}{\dot\phi}\delta\phi$ and  also simplifies the  perturbations equations (\ref{p1}), (\ref{p2}), (\ref{chi}) and (\ref{p5}) as following: 
\begin{eqnarray}
&\left(3H^2+3\frac{H\dot{F}}{F}-\frac{\kappa^2\omega\phi^2_0}{2F}\right)\Phi-\left(H+\frac{\dot{F}}{2F}\right)\frac{\Delta}{a^2}\chi=-\frac{1}{2F}\left\lbrace\left(\kappa^2\omega\dot\phi_0-3\frac{H\dot{F}}{\dot\phi_0}\right)\dot{\delta\phi}
\right.\nonumber
\\
& \left.
+\left[\frac{\dot{F}}{\dot\phi_0}\left(3H^2+3\dot{H}+\frac{\Delta}{a^2}\right)-3H\left(\frac{\dot F}{\dot\phi_0}\right)^.+\frac{1}{2}\left(\kappa^2\omega_{,\phi}\dot\phi_0^2+\kappa^2V_{,\phi}-\frac{f_{,\phi}}{2}\right)\right]\delta\phi\right\rbrace
\end{eqnarray}
 \begin{eqnarray}
 \left[\frac{\kappa^2\omega\dot\phi_0}{2F}-\frac{H\dot{F}}{2F\dot\phi_0}+\frac{1}{2F}\left(\frac{\dot{F}}{\dot{\phi_0}}\right)^.\right]\delta\phi+\frac{\dot{F}}{2F\dot{\phi_0}}\dot{\delta\phi}=\left(H+\frac{\dot F}{2F}\right)\Phi
 \end{eqnarray}
 \begin{eqnarray}
 \left(H+\frac{\dot{F}}{F}\right)\chi+\dot{\chi}=\frac{\dot{F}}{F\dot{\phi_0}}\delta\phi+\Phi
 \end{eqnarray}
 \begin{eqnarray}
 \begin{split}
&\ddot{\delta\phi}+\left(3H+\frac{\omega_{,\phi}}{\omega}\dot\phi_0\right)\dot{\delta\phi}+\left[-\frac{\Delta}{a^2}+\left(\frac{\omega_{,\phi}}{\omega}\right)_{,\phi}\frac{\dot{\phi^2}}{2}+\left(\frac{2V_(,\phi)-f_{,\phi}/\kappa^2}{2\omega}\right)_{,\phi}\right]\delta\phi=
\\
&\dot{\phi_0}\dot\Phi+\left(2\ddot{\phi_0}+6H\dot\phi_0+\frac{\omega_{,\phi}}{\omega}\phi_0^2\right)\Phi-\dot{\phi_0}\frac{\Delta}{a^2}\chi+\frac{F_{,\phi}}{2\omega\kappa^2}\delta R
\end{split}
\end{eqnarray} 
By using these equations and also relations (\ref{eq1}) and (\ref{sp}), we could derive two following equations in terms of $\delta\phi$ and $\Phi$:
\begin{equation}
\begin{split}
&\ddot{\delta\phi}+\left[3H+\frac{\omega_{,\phi}}{\omega}\dot{\phi_0}+\frac{\dot{\phi_0}+\frac{F_{,\phi}}{\kappa^2\omega}\left(\frac{\dot{F}}{F}-H\right)}{2FH+\dot{F}}\left(\kappa^2\omega\dot{\phi_0}-3\frac{\dot{F}H}{\dot\phi_0}\right)\right]\dot{\delta\phi}\\
&+\left\lbrace-\frac{\Delta}{a^2}+\frac{\dot\phi_0+\frac{F_{,\phi}}{\kappa^2\omega}\left(\frac{\dot{F}}{F}-H\right)}{2FH+\dot{F}}\left[\frac{\dot{F}}{\dot\phi}\left(3H^2+3\dot{H}+\frac{\Delta}{a^2}\right)-3H\left(\frac{\dot{F}}{\dot{\phi_0}}\right)^.-\kappa^2\omega(\ddot{\phi_0}+3H\dot{\phi_0})\right]
\right.
\\
& \left.
+\left(\frac{\omega_{,\phi}}{\omega}\right)_{,\phi}\frac{\dot\phi^2}{2}+\left(\frac{2V_{,\phi}-f_{,\phi}}{2\omega}\right)_{,\phi}\right\rbrace\delta\phi=\left(\dot{\phi_0}-3\frac{F_{,\phi}}{\kappa^2\omega}H\right)\dot\Phi+\left[\frac{\omega_{,\phi}}{\omega}\dot\phi_0^2
+2\ddot\phi_0+6H\dot{\phi_0}\right.
\\
& \left.
-\frac{F_{,\phi}}{\omega\kappa^2}(6\dot{H}+12H^2)-2F\frac{\dot{\phi_0}+\frac{F_{,\phi}}{\kappa^2\omega}\left(\frac{\dot{F}}{F}-H\right)}{2FH+\dot{F}}\left(3H^2+3H\frac{\dot{F}}{F}-\frac{\kappa^2\omega\dot\phi_0^2}{2F}\right)\right]\Phi
\end{split}
\end{equation}
\begin{eqnarray}
\Phi=\frac{1}{2FH+\dot{F}}\left(\kappa^2\omega\dot\phi_0-\frac{H\dot{F}}{\dot{\phi_0}}+\left(\frac{\dot{F}}{\dot{\phi_0}}\right)^.\right)\delta\phi+\frac{\dot{F}}{\dot{\phi_0}(2FH+\dot{F})}\dot{\delta\phi}
\end{eqnarray}
and then after some lengthy and tedious calculations we obtain the Mukhanov-Sasaki equation \cite{Hwang1}:
\begin{equation}
\begin{split}
&\ddot{\delta\phi}+\left\lbrace3H+\frac{(1+\dot{F}/2FH)^2}{\omega+3\dot{F}^2/2F\kappa^2\dot\phi_0^2}\left[\frac{\left(\omega+\frac{3\dot F^2}{2F\dot{\phi_0^2}\kappa^2}\right)}{\left( 1+\frac{\dot F}{2HF}\right)^2}\right]^.\right\rbrace\dot{\delta\phi}-\\
&\left\lbrace\frac{\Delta}{a^2}+\frac{H}{a^3\dot{\phi_0}}\frac{(1+\dot{F}/2FH)^2}{\omega+3\dot{F}^2/2F\kappa^2\dot\phi_0^2}\left[\frac{\left(\omega+\frac{3\dot F^2}{2F\dot{\phi_0^2}\kappa^2}\right)}{\left( 1+\frac{\dot F}{2HF}\right)^2}a^3\left(\frac{\dot{\phi_0}}{H}\right)^.\right]^.\right\rbrace\delta\phi=0
\end{split}
\end{equation}
Changing variable to $\delta\phi=-\frac{\dot{\phi_0}}{H}\tilde{\mathcal{R}}$ allows us to rewrite the above equation as:
\begin{align}
\label{MS}
\frac{\left( H+\frac{\dot F}{2F}\right)^2}{a^3\left(\omega\dot\phi_0^2+\frac{3\dot F^2}{2F\kappa^2}\right)}\left[\frac{a^3\left(\omega\dot\phi_0^2+\frac{3\dot F^2}{2F\kappa^2}\right)}{\left( H+\frac{\dot F}{2F}\right)^2}\dot{\tilde{\mathcal{R}}}\right]^.-\frac{1}{a^2}\Delta\tilde{\mathcal R}=0
\end{align}
Now in order to obtain the initial condition of this equation, we return to the Newtonian gauge. Substituting (\ref{Psi}) and (\ref{phi}) into (\ref{R}) and noting that at the beginning of inflation era, only the scalar field fluctuations contributes in $\tilde{\mathcal{R}}$, we find that:
\begin{align}\label{initial condition}
\tilde{\mathcal R}_q\to-\frac{H}{(2\pi)^{3/2}a(t)\sqrt{2qF\omega}\dot\phi_0}\exp\left(-iq\int_{t^*}^{t}\frac{dt'}{a(t')}\right)
\end{align}
where $\tilde{\mathcal R}_q$ is the Fourier transform of $\tilde{\mathcal R}$.  Beyond the horizon where  $q/a\ll H$, equation (\ref{MS}) has two growing and decaying perturbation modes \cite{Hwang1}. The growing mode  is a non-zero constant which is exactly the same as obtained in general relativity.
 
For any given function $f(R,\phi)$, one can solve equation (\ref{MS}) with the initial condition (\ref{initial condition}) which shows that, the curvature perturbation at sufficiently very early times, is independent of the behavior of the scalar potential. This means that the constant curvature perturbation, outside the horizon, depends only to the scalar potential evaluated when the perturbation mode leaves the horizon.  This feature is also similar to what is obtained in general relativity \cite{Weinberg} \\
\section{Tensor perturbations in generalized scalar-tensor gravity}
In this section we would like to consider the initial condition for tensor metric perturbation. It  plays an important role in cosmology since the quantum tensor fluctuation of metric generates the gravitational wave fluctuation which can be also observed. The tensor metric perturbations is a gauge invariant quantity and the initial condition problem for it 
can be treated in the same way as it is done for the scalar perturbations in the previous section.

Varying the action (\ref{action1}) with respect to the metric (\ref{metric}), considering only the tensor perturbation, gives the equation of motion of $\mathcal{D}_{ij}$ as following  \cite{living}
\begin{align}
\ddot{\mathcal{D}}_{ij}+\frac{(a^3F)^.}{a^3F}\dot{\mathcal{D}}_{ij}-\frac{\nabla}{a^2}\mathcal{D}_{ij}=0
\end{align}
Since for a single scalar field, the tenor component of the anisotropic inertia is zero, the scalar field is explicitly appeared in  the above equation through function $F$. It is straightforward to see that the tensor fluctuations have plane-wave solutions of the form $e_{ij}\mathcal{D}_q(t)e^{i\textbf q.\textbf x}$
where $\mathcal{D}_q(t)$  satisfies
\begin{align}\label{tensor}
\ddot{\mathcal{D}}_q+\frac{(a^3F)^.}{a^3F}\dot{\mathcal{D}}_q+\frac{q^2}{a^2}\mathcal{D}_q=0
\end{align}
Just like the scalar perturbations, at sufficiently very early times where $q/a\gg H$,  the equation (\ref{tensor}) can be solved using the WKB approximation
\begin{align}\label{TS}
\mathcal{D}_q(t)\to z(t)\exp{\left(-iq\int_{t^*}^t\frac{dt'}{a(t')}\right)}
\end{align}
in which $z(t)$ is a slowly varying function in comparison with $q/a$. Substituting (\ref{TS}) into (\ref{tensor}), up to leading order in $q/a$ we get
\begin{align}
\dot{z}+\frac{1}{2}\left(\frac{(a^3F)^{.}}{a^3F}-H\right)z=0
\end{align}
Therefore
\begin{align}\label{53}
z(t)=\frac{K}{a\sqrt{F}}
\end{align}
 where $K$ is an arbitrary constant of integration whose $q$-dependence must be determined. Following \cite{Weinberg}, by comparing (\ref{53}) and (\ref{eq12}), we choose $K=\frac{\kappa}{(2\pi)^{3/2}\sqrt{2q}}$. Thus the initial condition of tensor fluctuation for $a\longrightarrow 0$ is
 \begin{align}\label{54}
\mathcal{D}_q(t)\to\frac{\kappa}{(2\pi)^{3/2}a\sqrt{2qF}}\exp{\left(-iq\int_{t^*}^t\frac{dt'}{a(t')}\right)}
\end{align}
Thus the general solution of (\ref{tensor}), satisfying the conditions (\ref{condition}) would be
\begin{align}
\mathcal{D}_{ij}(\textbf{x},t)=\sum_{\lambda=+,\times}\int d^3q\left[\mathcal{D}_q(t)e^{i\textbf{q}.\textbf{x}}\beta(\textbf{q},\lambda)\hat{e}_{{ij}}(\hat{q},\lambda)+\mathcal{D}^*_q(t)e^{-i\textbf{q}.\textbf{x}}\beta^*(\textbf{q},\lambda)\hat{e}_{ij}^*(\hat{q},\lambda)\right]
\end{align}
where $\mathcal{D}_q(t)$ is given by (\ref{54}) determined by the evolution
of fluctuations in the early universe. As mentioned before, at very early universe, the fluctuations are described by quantum fields. This means that,  in the above relation $\beta(q,\lambda)$ and $\beta^*(q,\lambda)$ are the annihilation and creation operator for graviton and satisfy the standard commutation relations 
\begin{align}
\left[\beta(\textbf{q},\lambda),\beta(\textbf{q}', \lambda')\right]=0,\hspace{0,5in}\left[\beta(\textbf{q},\lambda),\beta^*(\textbf{q}',\lambda')\right]=\delta^3(\textbf{q}-\textbf{q}')\delta_{\lambda\lambda'}
\end{align}
In the limit of $q/a\ll H$, the equation (\ref{tensor}) has two independent solutions which one of them is a non-zero constant while the other one is decaying \cite{Hwang1}. 

\section{Conclusion}
The evolution of cosmological fluctuations in radiation dominated universe depends on the initial conditions through two scalar and tensor gauge invariant quantities, $\mathcal{R}_q^0$ and $\mathcal{D}_q^0$ respectively. These are the conserved curvature perturbation and gravitational wave amplitude outside the horizon respectively. To find these, it is required to go back to the inflationary era and obtain the equations governing the scalar and tensor perturbations. The initial conditions of these equations come from the very early stages of inflation when the fluctuations are indeed quantum fields. Furthermore, in general relativity \cite{Weinberg}, at very early times of inflation when the fluctuations are inside the horizon, the behavior of the scalar and tensor fluctuations are independent of the inflaton's potential. While outside the horizon, the constant quantities $\mathcal{R}_q^0$ and $\mathcal{D}_q^0$, will be depended on the nature of the potential near the horizon crossing. For studying the fluctuations inside the horizon, we need only solve the equations governing the gauge invariant quantities with their initial conditions; without any arbitrary assumptions about the strength of cosmological fluctuations \cite{Weinberg}.\\
In the present paper, we have studied in more details, the points explained above in the context of generalized scalar tensor gravity. We have discussed the initial conditions of the scalar and tensor perturbations in the generalized gravity. The equations of the scalar perturbations have been derived in the Newtonian gauge  and then their solution have been obtained at very early times using WKB approximation. It has been shown that at very early times, we could ignore the scalar metric perturbations with respect to scalar field fluctuation just as general relativity. To find the initial condition, we have quantized the scalar and tensor fluctuations. \\
In order to achieve the equation governing the evolution of curvature perturbation, we have introduced a new gauge in which $\Psi=0$ and furthermore, the fluctuations of the scalar field is related to the fluctuations of the arbitrary function $F$ via  $\delta F/\dot{F}=\delta\phi/\dot{\phi_0}$. Then, in this gauge, we have derived the generalized Mukhanov-Sasaki equation  using the equations, governing the evolution of the scalar perturbations and doing some straightforward algebra. Then, using the initial condition already obtained for the scalar perturbations, we could set the initial condition of the curvature perturbation. Also in generalized scalar-tensor gravity, the equation of curvature perturbation has two independent adiabatic solutions outside the horizon. One is a non-zero constant and the other presents a decaying mode just as we have seen in general relativity. This non-zero constant curvature perturbation allows us to connect the past distant to the nearly present times.\\
Since the tensor fluctuation of metric is the origin of the gravitational wave and can be interpreted as a spin-2 particle (graviton), we have studied the tensor perturbation of metric and obtained the initial condition of it and then quantized it in the context of $f(R,\phi)$ gravity.\\
\section*{Acknowledgements}
This work has been supported by a grant from university of Tehran.

\end{document}